\RequirePackage{lineno}
\documentclass[twocolumn,twoside,slac_two]{revtex4}
\usepackage{graphicx}
\usepackage{fancyhdr}
\usepackage{epstopdf}
\begin{document}
%\linenumbers
\title{Evolution of structures in two particle correlations in RHIC Cu+Cu 
collisions as a function of centrality and momentum}


\author{L.C. De Silva for the STAR collaboration}
\affiliation{Department of Physics and Astronomy, Wayne State University, Detroit, MI 48201, USA}

\begin{abstract}
Two particle correlation measurements in heavy ion collisions at RHIC have shown an extended near side correlation in $\Delta\eta$ relative to p+p for both, momentum triggered and untriggered analyses. This phenomenon is also known as the "ridge". An investigation into the momentum and centrality dependence of two particle correlations is presented for Cu+Cu 200 GeV collisions from the STAR experiment. We extract the amplitude, $\Delta\eta$ and $\Delta\phi$ widths from the near side correlation structure, and show how its parameters depend on centrality and the lower transverse momentum cut-off. Implications for the origin of the ridge will be discussed.
\end{abstract}

\maketitle

\thispagestyle{fancy}


\begin{figure}[h]
\centering
\includegraphics[width=0.3\textwidth]{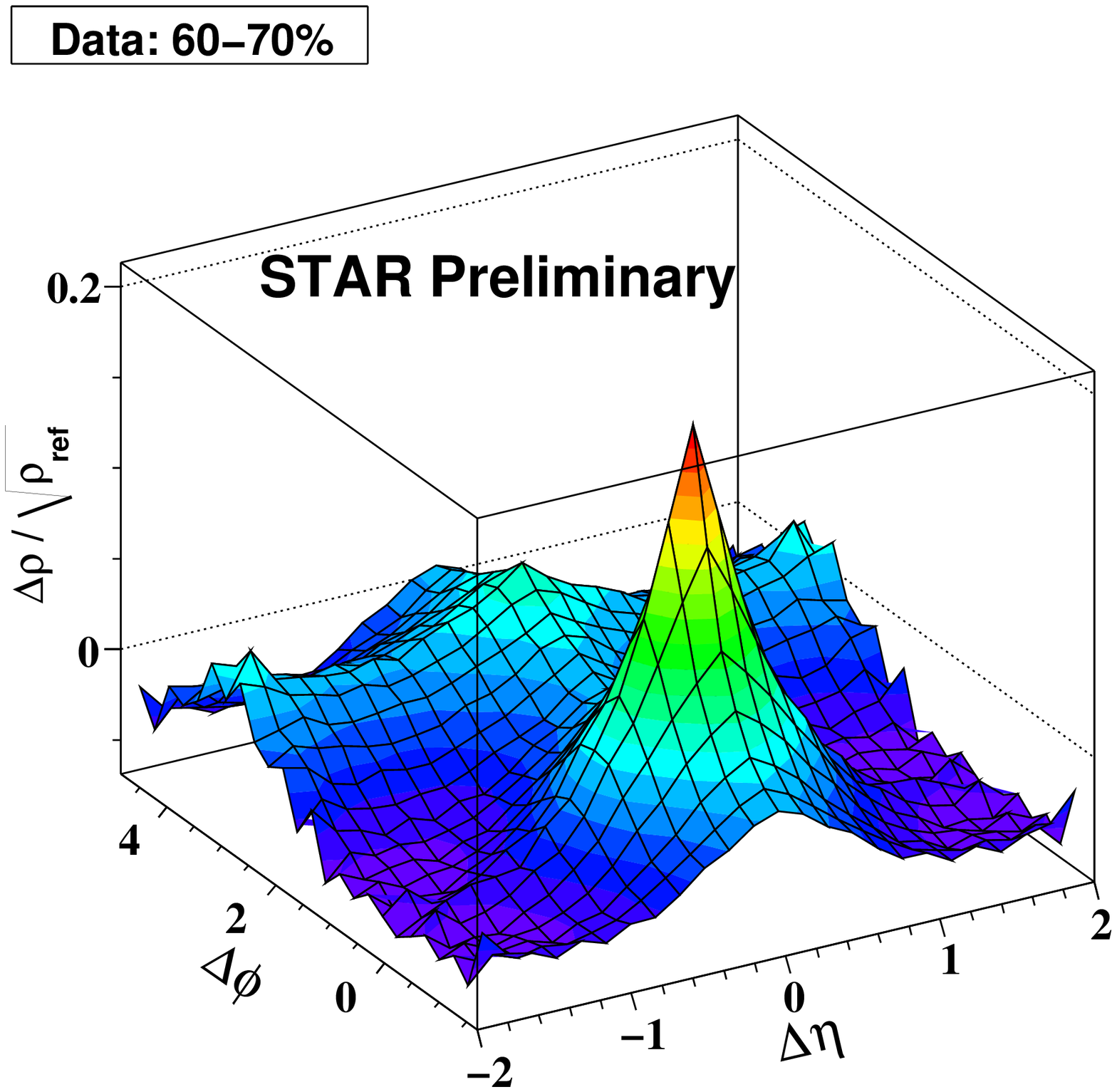}
\includegraphics[width=0.3\textwidth]{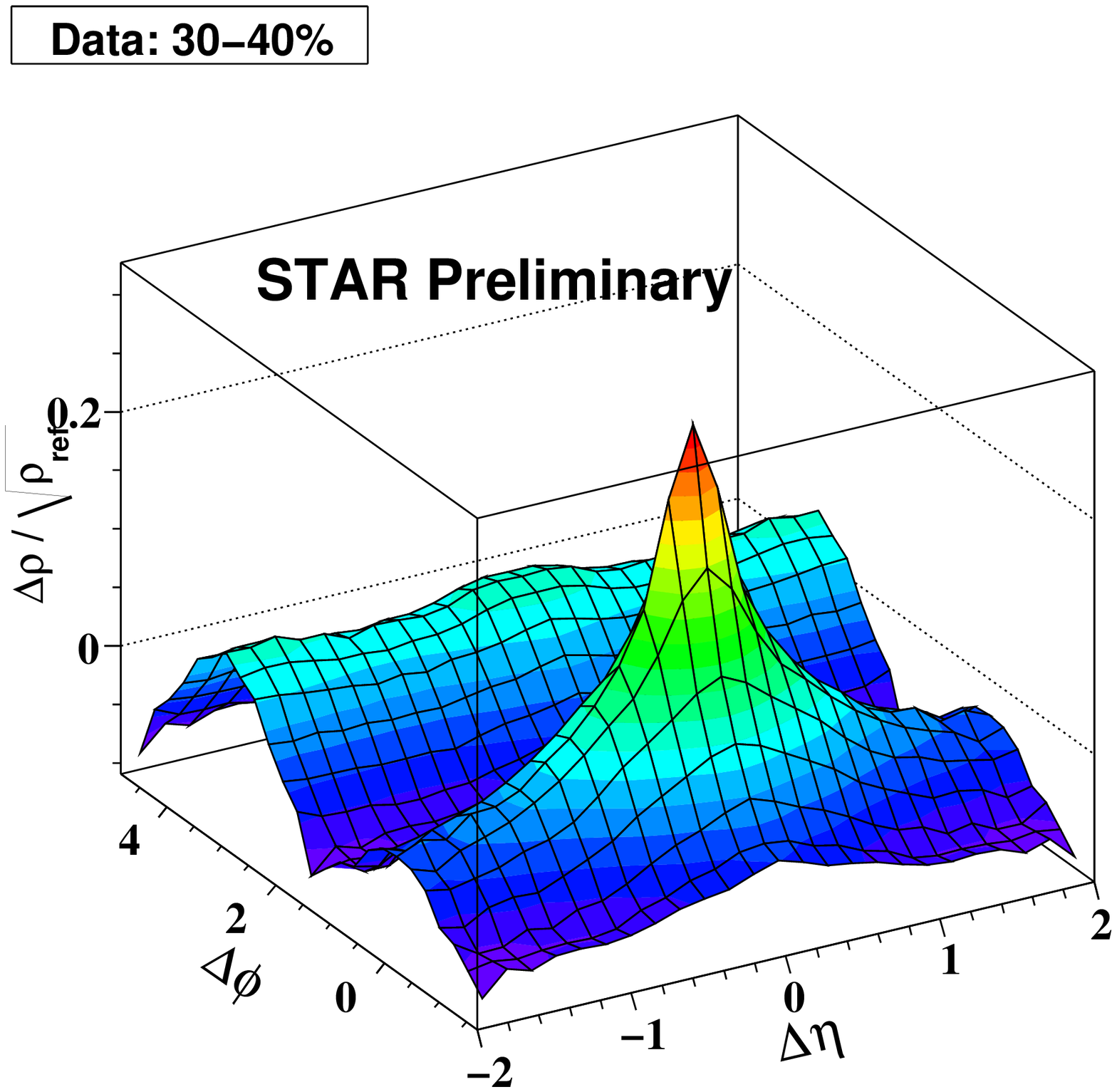}
\includegraphics[width=0.3\textwidth]{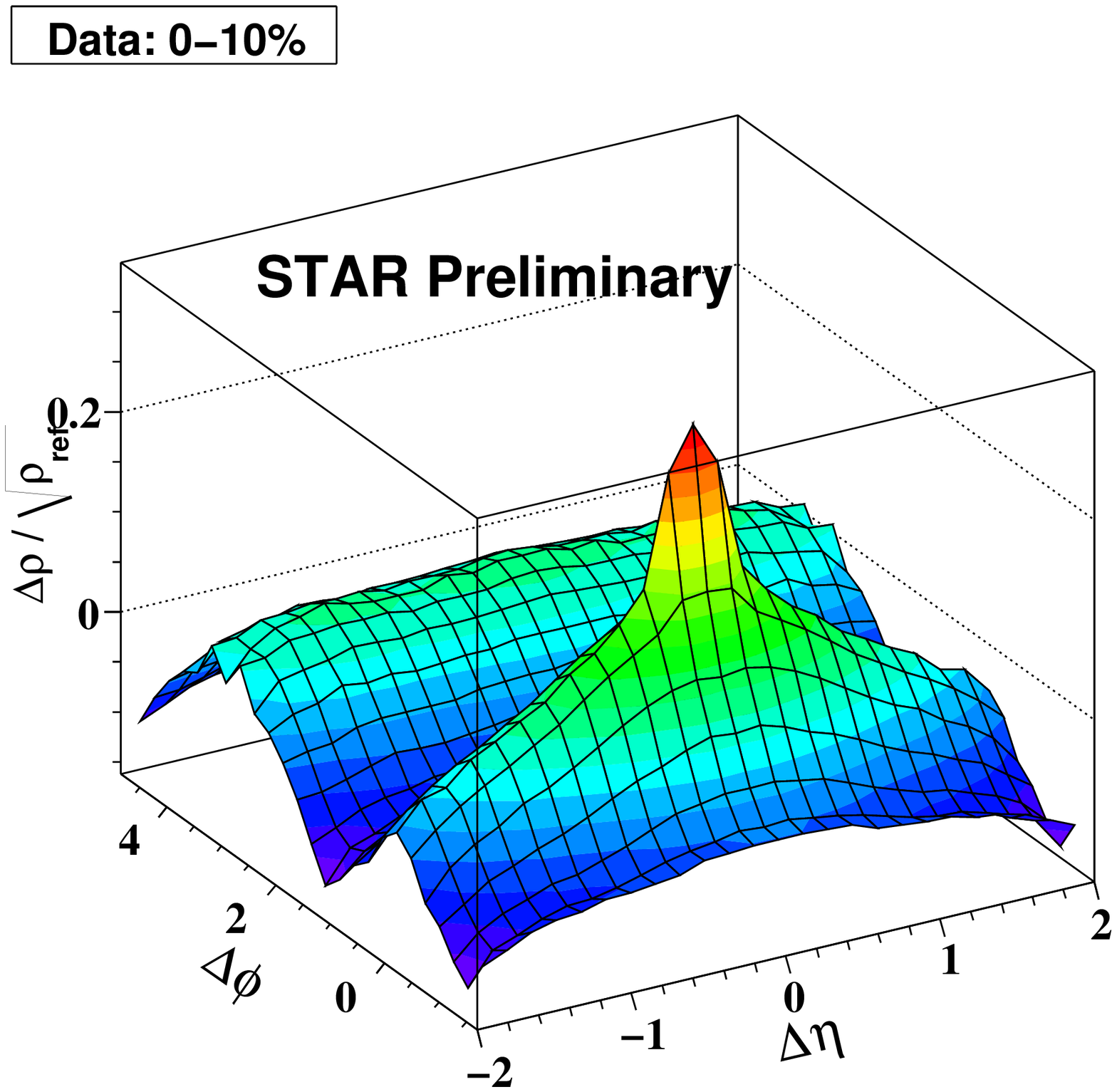}
\caption{Evolution of the raw normalized correlation structure for 60-70\%, 30-40\%, and 0-10\% centrality, respectively.}
\label{figure:raw}
\end{figure}

%%%%%%%%%%%%%%%%%%%%%%%%%%%%%%%%%%
\section{Introduction}

High momentum two-particle correlation studies with the STAR experiment at RHIC have revealed
a long-range correlation structure in $\Delta\eta$ associated with a near-side trigger particle
($|\Delta\phi|$ $<$ $\pi$/2) \cite{james, putschke, abelev}. This structure is known as the ÒridgeÓ \cite{putschke}. It was initially observed for 4 GeV/c $ >$ p$_{T,trig}$ $>$ 3 GeV/c and p$_{T,asso} $ $>$ 2 GeV/c, but persists to even higher trigger p$_{T}$ ($>$ 6 GeV/c), indicating that it is in fact associated with jet production.

A similar ridge-like structure is also observed in untriggered two particle analyses, where all possible particle pairs contribute to the two-particle correlation structure \cite{daugherity, adames}. This is not easily explained by jet modification, though, because the vast majority of particles is well below the pQCD limit where fragmentation should be applicable \cite{albino}. Various theoretical approaches have been suggested ranging from mini-jet production \cite{trainor}  to parton momentum kick models \cite{wong}  to the radial expansion of initial conditions of the bulk matter \cite{voloshin, gavin}. 

We use the analysis procedure for the untriggered correlation functions developed by STAR to compare the centrality dependence in the 200 GeV Cu+Cu collisions to the 200 GeV Au+Au system studied in \cite{daugherity}. We further investigate the evolution of the ridge structure from the untriggered analysis to a momentum triggered analysis by raising the lower momentum cut of both correlated particles. 
In the following section we briefly describe some of the details of the analysis and then show results for the centrality and momentum dependence. 

%%%%%%%%%%%%%%%%%%%%%%%%%%%%%%%%%%
\section{Analysis procedure}

We use charged particle tracks detected in the STAR Time Projection Chamber with a tracking imposed minimum  p$_{T}$ $>$ 0.15 GeV/c, $|$ $\eta$ $|$ $<$ 1 and full 2$\pi$ azimuth. 7 Million minimum bias triggered 200 GeV Cu+Cu events were analyzed. Track finding efficiency corrections have not been applied, but are estimated to yield a 10 -20\% correction for the extracted amplitude of the correlation distribution. The analysis is performed by determining PearsonÕs correlation coefficient, which is defined as the ratio of the co-variance over the product of the standard deviations of two variables. This is also known as an auto-correlation. It can be expressed as $\Delta\rho$ / $\sqrt{\rho_{ref}}$. The ratio is obtained by calculating particle pair densities  and using the event-mixing technique to subtract the uncorrelated background. More specifically we construct $\rho_{sib}$, the sibling pairs from particles within the same event, and $\rho_{ref}$, the uncorrelated reference pairs, by mixing particles from independent events in order to calculate  $\Delta\rho$ = $\rho_{sib}$ $-$ $\rho_{mix}$. The pair densities were measured as number of pairs per unit area in relative angle bins ($\Delta\eta$ = $\eta_{1}$ $-$ $\eta_{2}$, $\Delta\phi$ = $\phi_{1}$ $-$ $\phi_{2}$). Hence our notation of the normalized correlation coefficient gives the number of correlated pairs per trigger particle. 

Fig.~\ref{figure:raw} shows the normalized correlation distribution for three centralities in 200 GeV Cu+Cu events. These histograms are fit with a multi-dimensional fit function with 11 free parameters, some of which could be constrained by additional independent measurements in STAR, such as the elliptic flow, the HBT correlation strength and the e$^{+}$e$^{-}$ annihilation cross section. Further details can be found in \cite{daugherity, adames, trainor}. The remaining structures are directly related to the occurrence of a same side ridge
and an offsetting momentum conservation component on the away-side. The same-side structure in the untriggered analysis is approximated by a two dimensional Gaussian fit, whereas the away-side component is described by a cos$\Delta\phi$ function. 

%%%%%%%%%%%%%%%%%%%%%%%%%%%%%%%%%%
\section{Results}

\begin{figure}[t]
\centering
\includegraphics [width=0.3\textwidth]{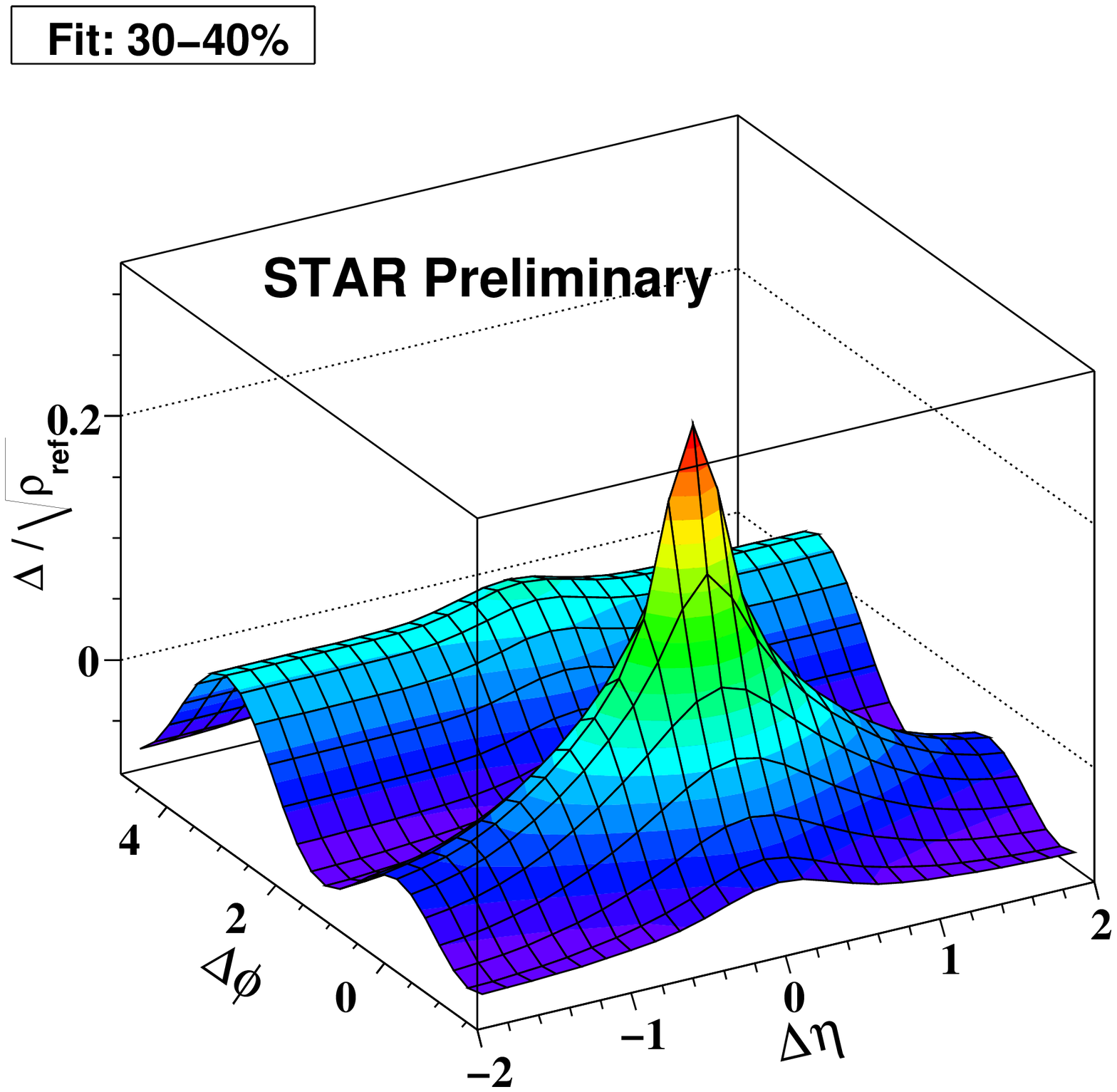}
\includegraphics [width=0.3\textwidth]{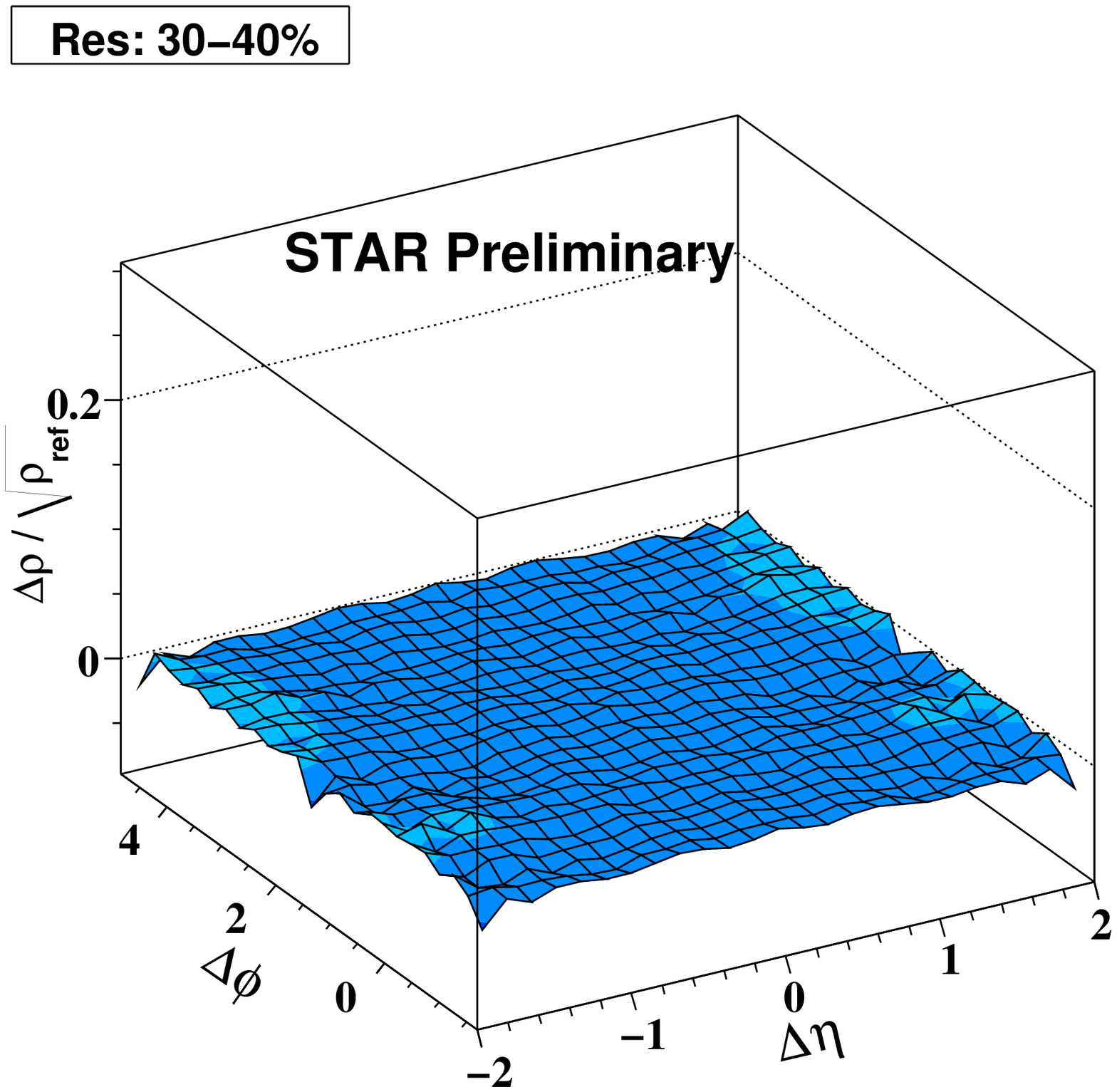}
\caption{Top: fit for 30-40\% centrality correlation function. Bottom: residuals (data-fit) for 30-40\% centrality correlation function.}
\label{figure:fit}
\end{figure}

Fig.~\ref{figure:fit} shows the fit function result (top) and the residual (data-fit) (bottom) for the 30-40 \% centrality bin. The fit has a $\chi^{2}$/n of 2.5, where $n$ is the number of degrees of freedom. For all fits $n=133$.

For all centralities the $\chi^{2}$/n varies between 2 and 4.5. In order to estimate the systematic error for a given fit parameter, $p$, we calculate $\chi^2/n$ distribution as a function of the parameter in question. For a given $\chi^2$ and $p$, all other parameters are adjusted to give the smallest value of $\chi^2$. We then locate the $\chi^2/n$ corresponding to the minimum $\chi^2/n$+1, and the difference in the parameter values for the minimum $\chi^2$ and  minimum $\chi^2$+1 ($p(\chi_{min}^2/n)$-$p(\chi_{min}^2/n +1)$) is interpreted as the uncertainty in that parameter.

\begin{figure*}[t]
\centering
\includegraphics[width=0.3\textwidth]{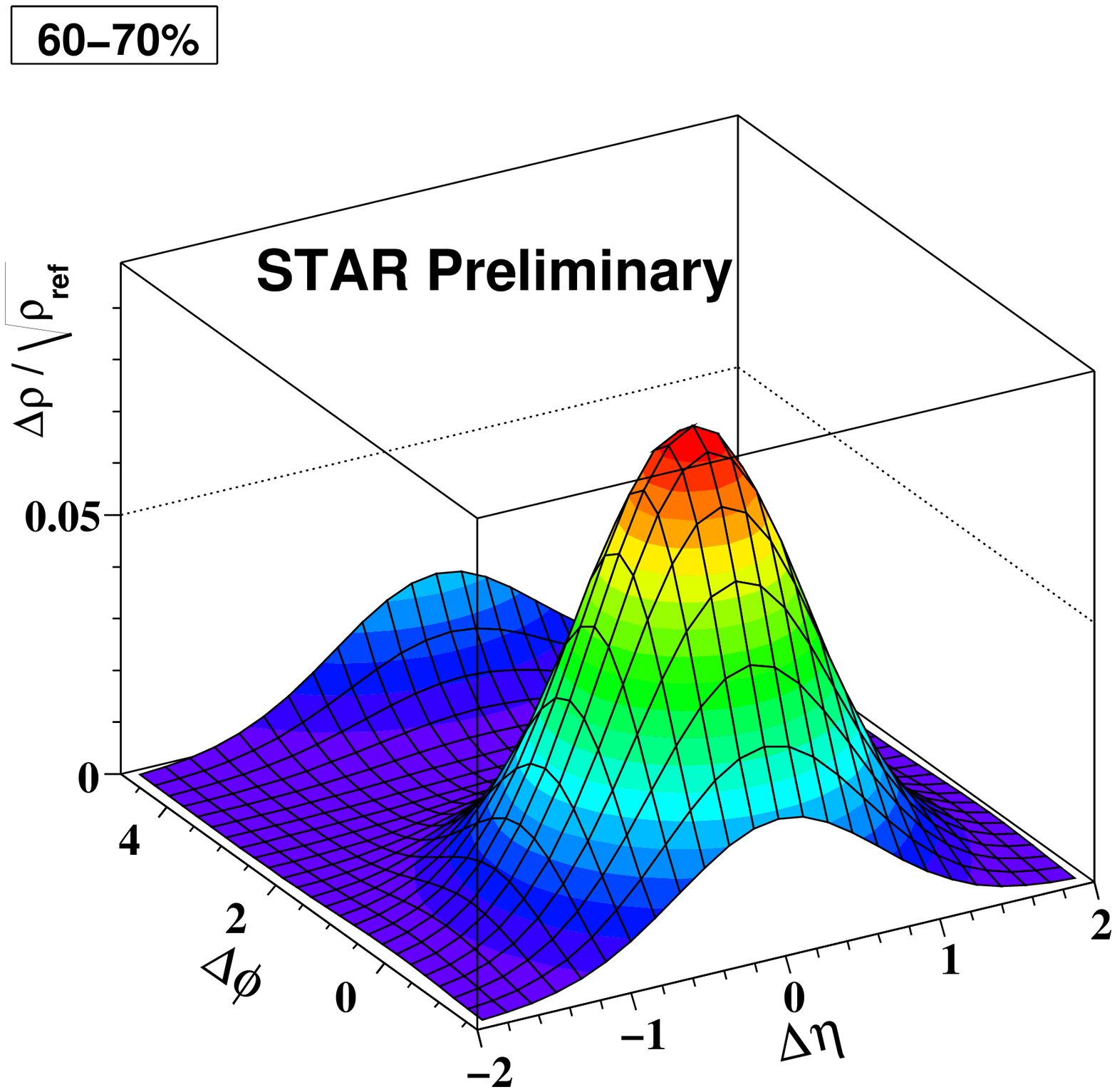}
\includegraphics[width=0.3\textwidth]{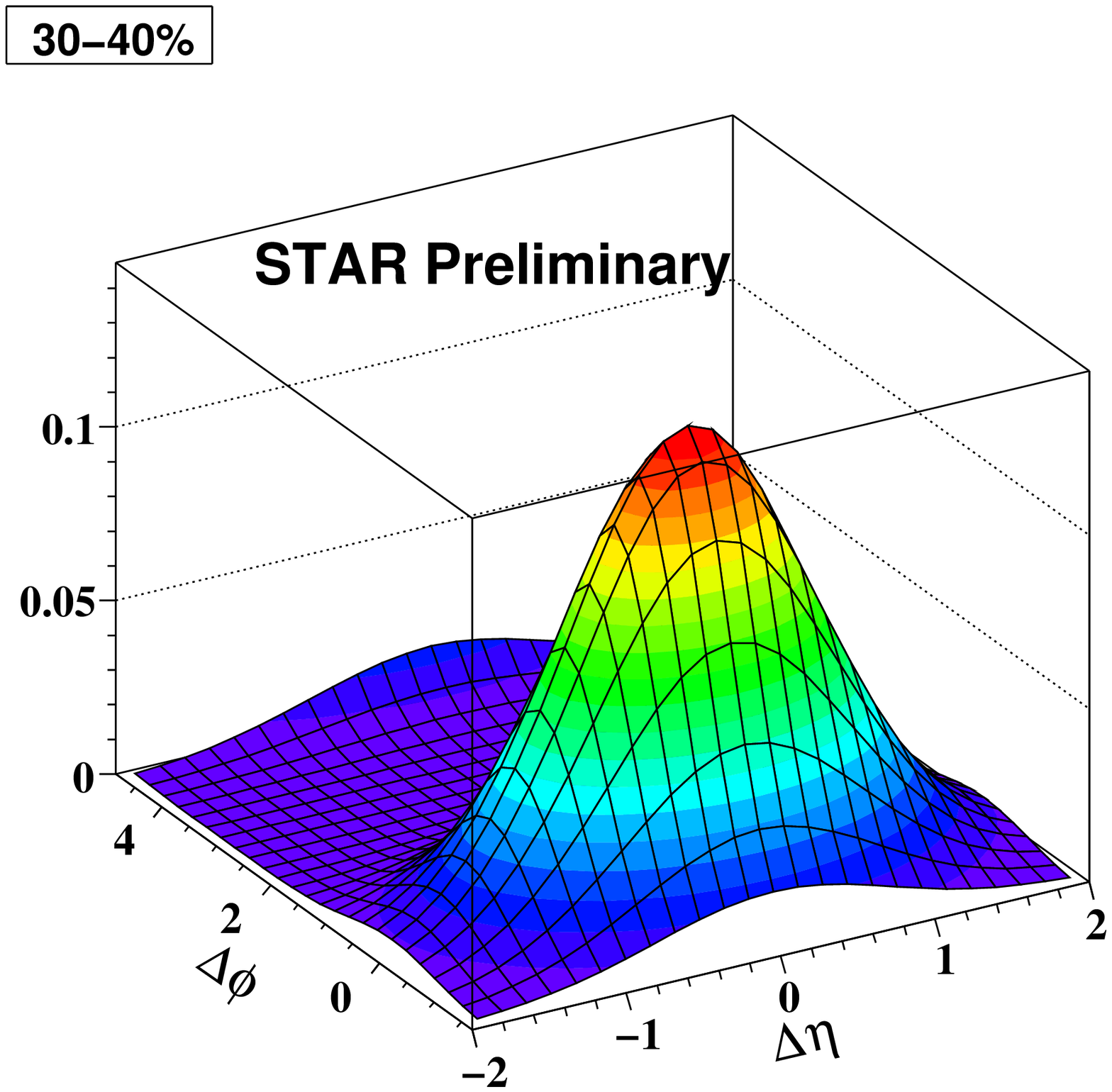}
\includegraphics[width=0.3\textwidth]{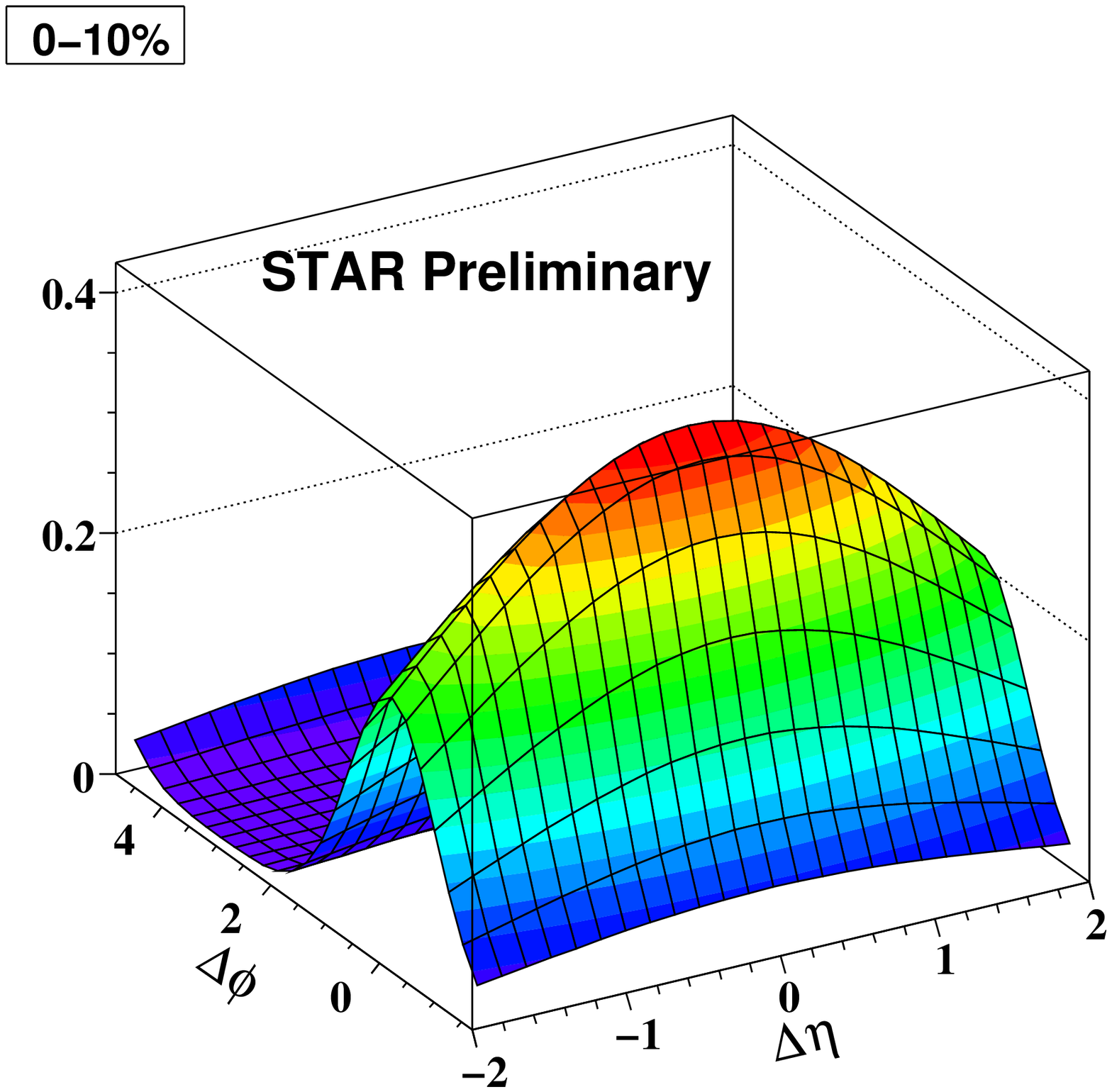}
\caption{Evolution of the decomposed 2d-Gaussian correlation structure for 60-70\%, 30-40\%, and 0-10\% centrality, respectively.}
\label{figure:2d}
\end{figure*}

\begin{figure*}[t]
\centering
\includegraphics[width=1.0\textwidth]{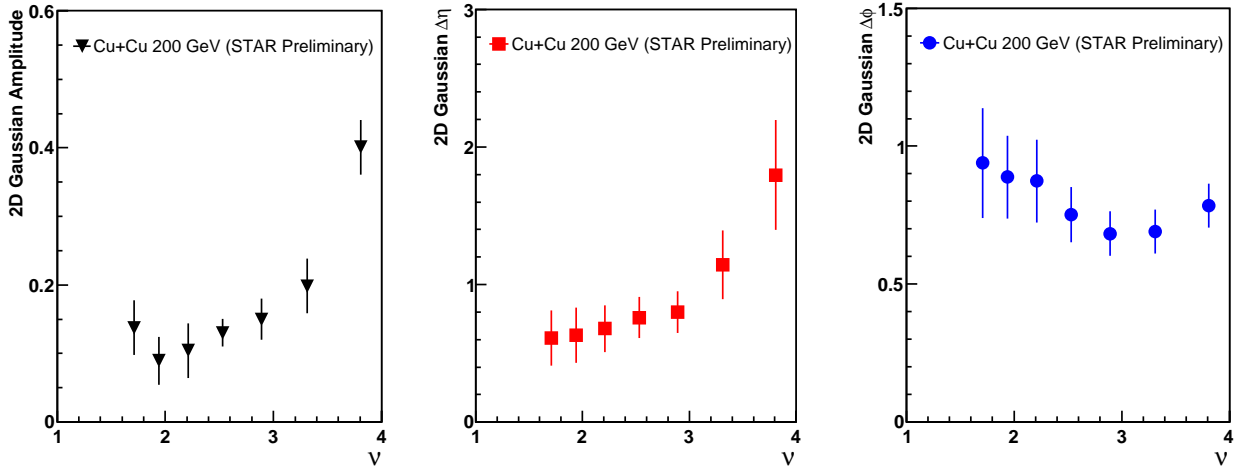}
\caption{Evolution of the parameters of the decomposed 2d-Gaussian correlation structure as a function of the path length variable $\nu$ = 2 $<$N$_{bin}$$>$/$<$N$_{part}$$>$. The plots show the amplitude, the $\Delta\eta$ width and the $\Delta\phi$ width,  respectively. Errors shown are systematic.}
\label{figure:parameters}
\end{figure*}

The 2d Gaussian fit to the same side structure is defined by three parameters: the amplitude, the $\Delta\eta$ width and the $\Delta\phi$ width. Fig.~\ref{figure:2d} shows the evolution of this specific structure as a function of centrality, Fig.~\ref{figure:parameters} shows the parameter evolution as a function of the path length variable $\nu$ = 2 $<$N$_{bin}$$>$/$<$N$_{part}$$>$, which should be compared to similar measurements for the Au+Au system \cite{daugherity}. 

\begin{figure*}[t]
\centering
\includegraphics[width=0.3\textwidth]{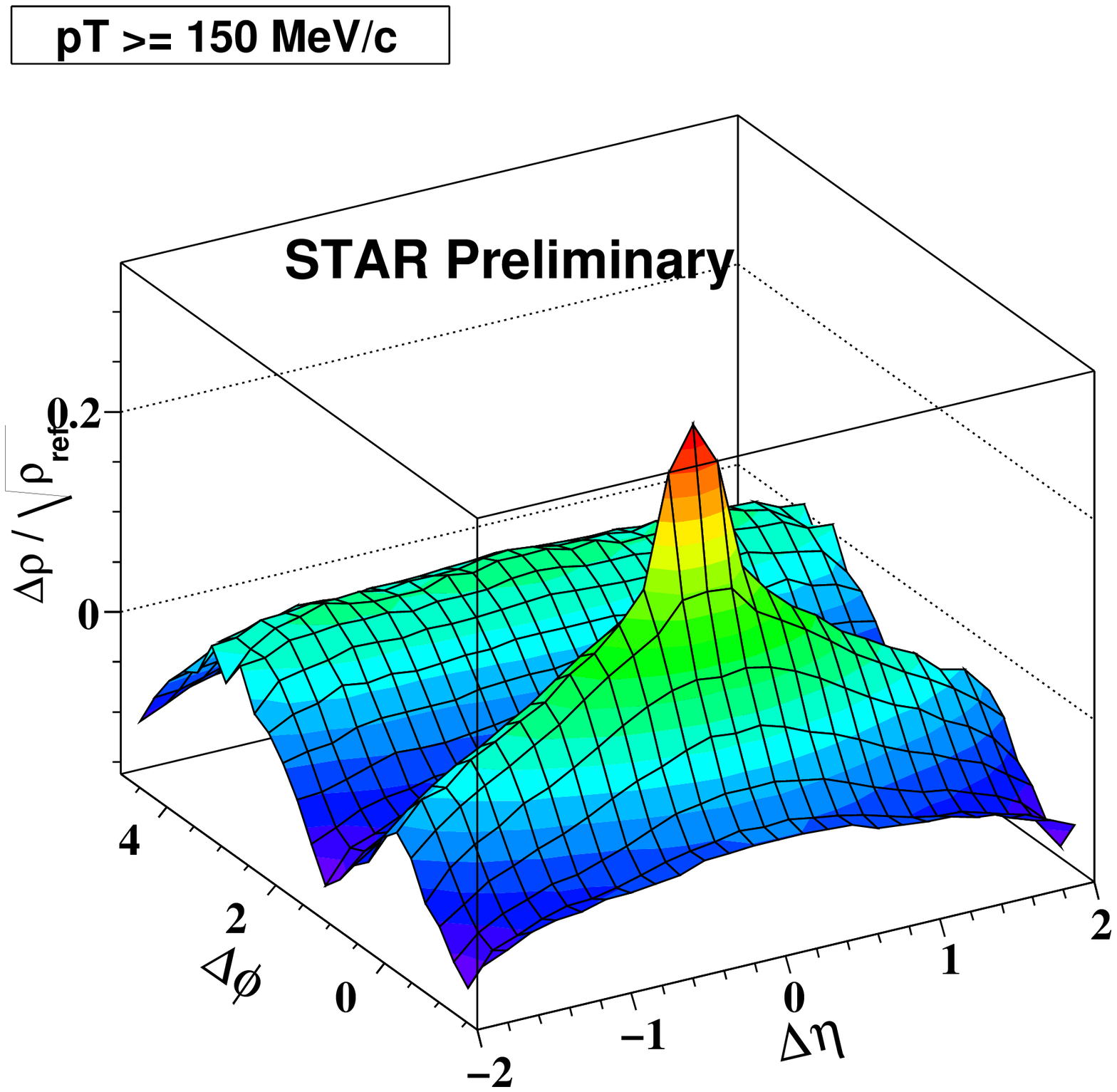}
\includegraphics[width=0.3\textwidth]{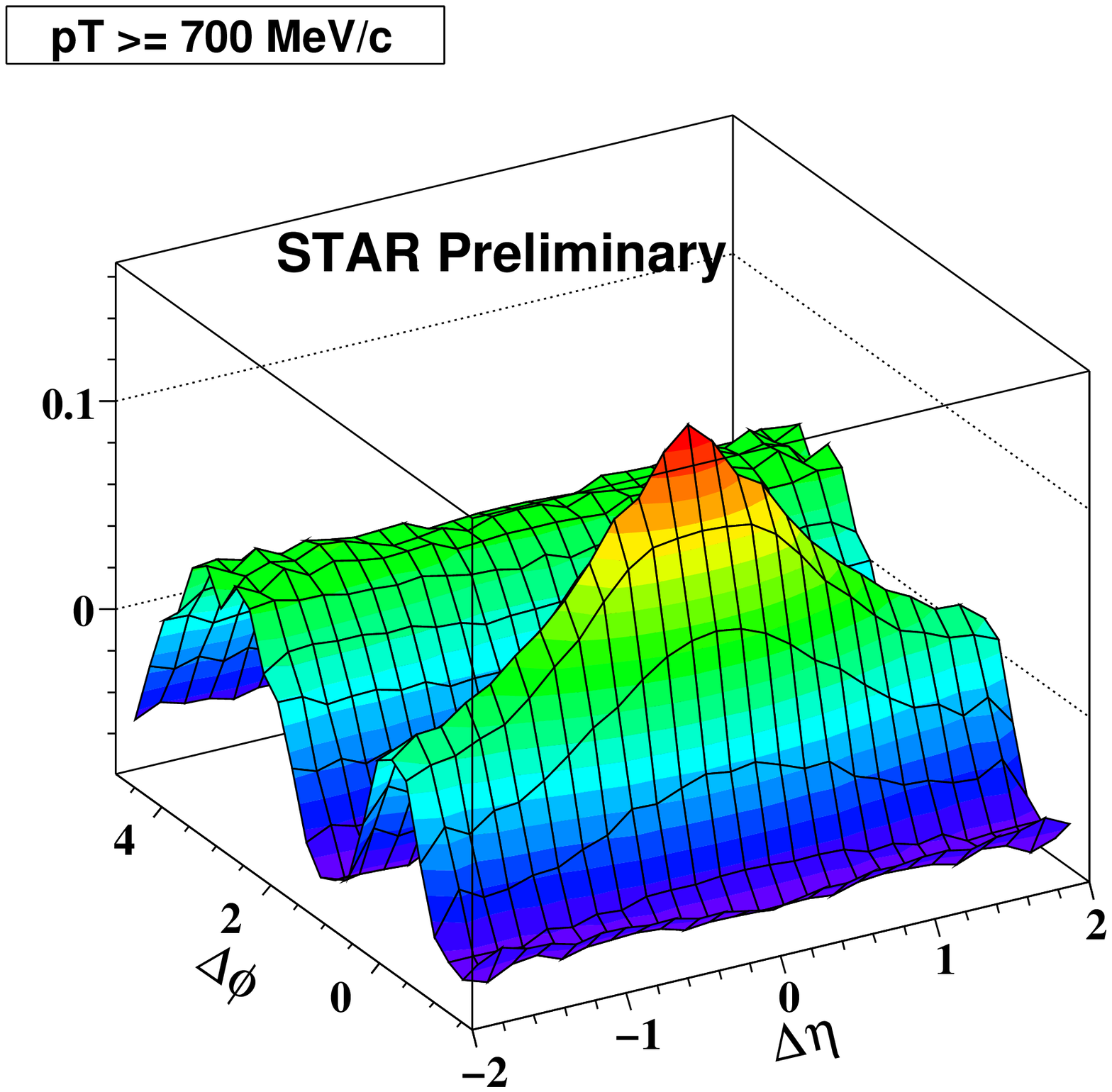}
\includegraphics[width=0.3\textwidth]{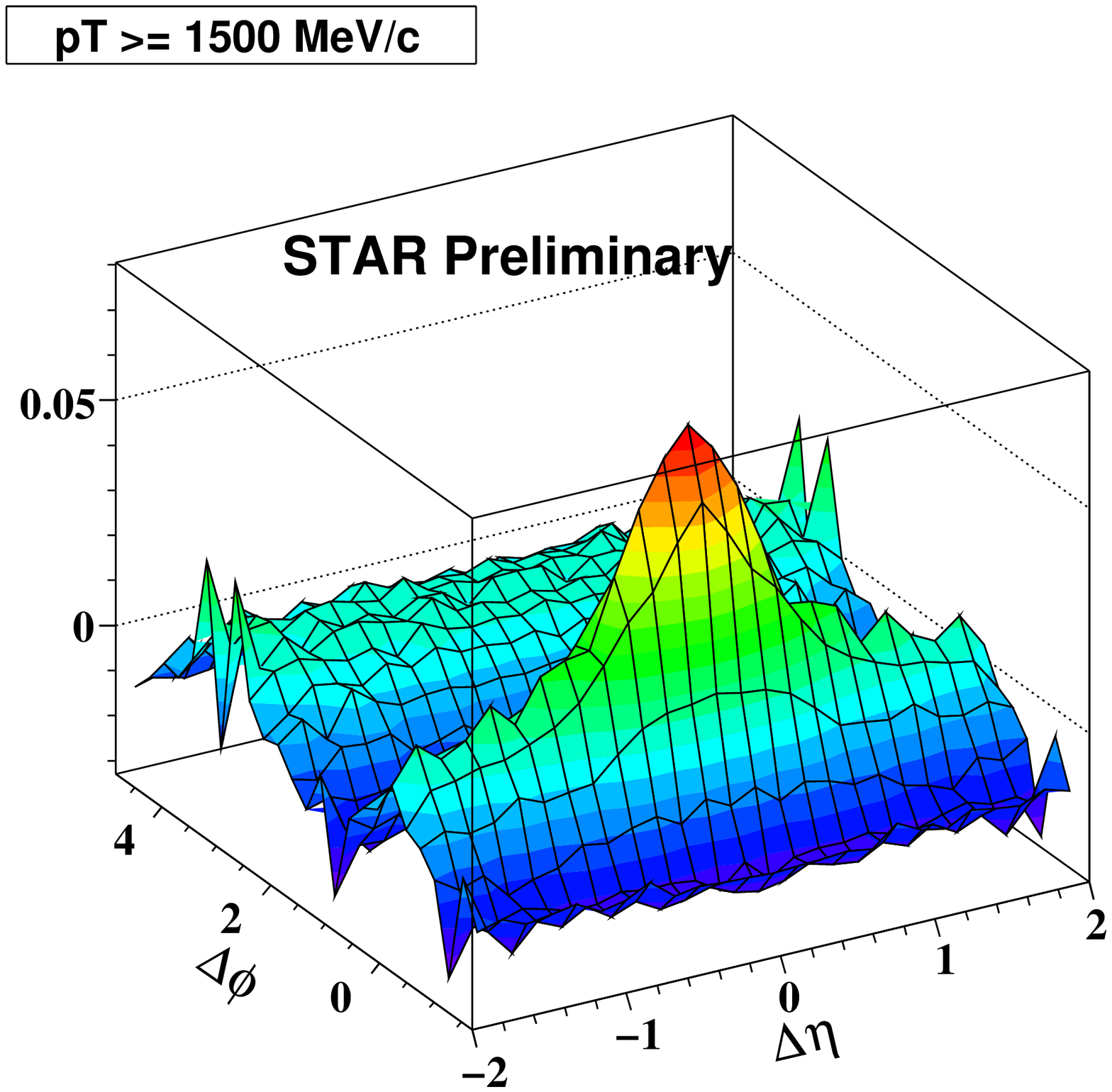}
\caption{Evolution of correlation structures as a function of transverse momentum for a lower p$_{T}$-cutoff of 150, 700, 1500 MeV/c, respectively. Please note the change in the correlation strength.}
\label{figure:pt}
\end{figure*}

A sizable increase in the same-side structure strength as a function of centrality is observed and consistent with the previous measurements in Au+Au. The significant abundance of correlated particles in a p$_{T}$ independent analysis, which is dominated by low p$_{T}$ (bulk) particles, is surprising and gave rise to many recent theoretical attempts to describe the data. Furthermore, as this structure exhibits similar features to the ridge in p$_{T}$ triggered analysis, the question arises whether the origin of the near-side structures is common for both bulk and jet dominated correlation distributions. 
In order to address this question experimentally we have imposed an additional lower p$_{T}$ cut on the accepted particles to our correlation analysis and gradually raised it (in 200 MeV/c steps from 0.15-1.5 GeV/c) for both particles in our correlation functions. Fig.~\ref{figure:pt} shows the change in the correlation function for three selected momentum thresholds. 

\clearpage
\begin{figure}[t]
\includegraphics [width=0.41\textwidth]{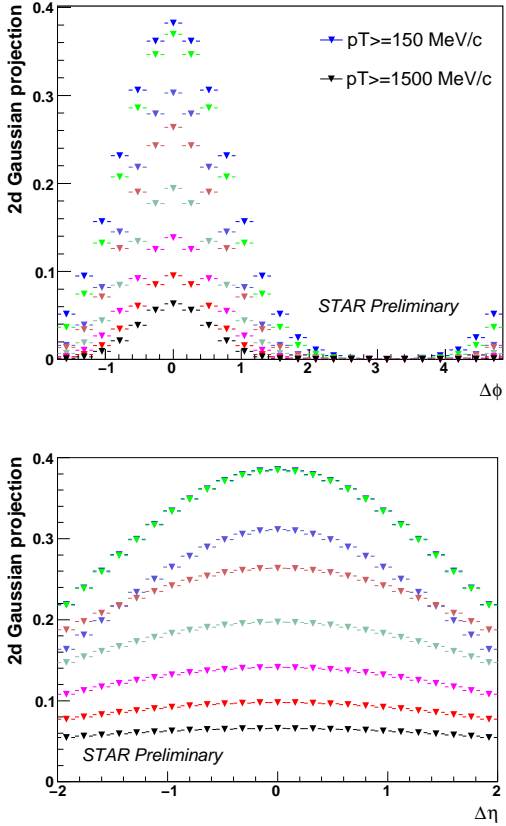}
\caption{$\Delta\phi$ (top) and $\Delta\eta$ (bottom) width as a function of the lower momentum cut-off. Curves are for p$_{T}$ thresholds of 150 MeV/c, 300 MeV/c, 500 MeV/c, 700 MeV/c, 900 MeV/c, 1100 MeV/c, 1300 MeV/c and 1500 MeV/c, respectively.}
\label{figure:deta-dphi}
\end{figure}
\begin{figure}[t]
\centering
\includegraphics [width=0.41\textwidth]{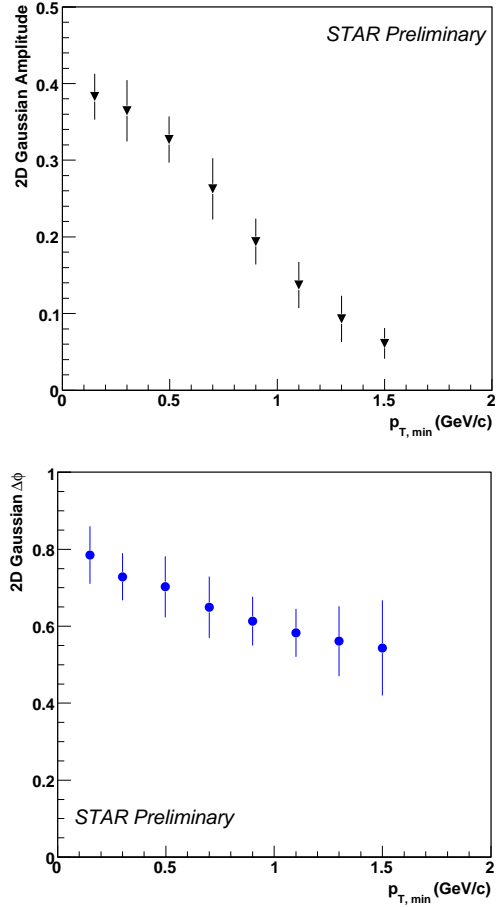}
\caption{Amplitude (top) and $\Delta\phi$-width (bottom) of the near-side 2d-Gaussian structure as a function of the lower momentum cut-off in the correlation structure. Errors shown are systematic.}
\label{figure:amp-phi}
\end{figure}
%

%\clearpage
%
%There is a smooth evolution from the wide Gaussian in central untriggered collisions to an even wider Gaussian at high p$_{T}$. The evolution is accompanied by the emergence of a jet-like structure at (0,0) in $\Delta\eta$ and $\Delta\phi$ for the higher momentum thresholds. Here we first attempt to fit the same-side distribution with a single 2d Gaussian out to 1.5 GeV/c, which is still dominated by a wider ridge. The resulting $\Delta\eta$ and $\Delta\phi$ momentum dependence is shown in Fig.~\ref{figure:deta-dphi}.
%
%From Fig.~\ref{figure:pt} we can deduce that replacing the single 2d Gaussian fit with a double Gaussian fit, as implied in the triggered ridge studies \cite{abelev} is required to further optimize the distinguishability between true jet and non-jet components. This work is in progress but it is beyond the scope of these proceedings.

There is a smooth evolution from the wide Gaussian in central untriggered collisions to an even wider Gaussian at high p$_{T}$ which is accompanied by the emergence of a jet-like structure at (0,0) in $\Delta\eta$-$\Delta\phi$ for the higher momentum thresholds. Although this narrow structure begins to impact the overall $\chi^{2}$/dof, the distribution is still dominated by the wider ridge in $\Delta\eta$, and thus we first attempt to fit the same-side distribution with the single 2d Gaussian out to 1.5 GeV/c. The resulting $\Delta\eta$ and $\Delta\phi$ momentum dependence is shown in Fig.~\ref{figure:deta-dphi}.

From Fig.~\ref{figure:pt}c we can deduce that replacing the single Gaussian with a double Gaussian fit, as implied in the triggered ridge studies \cite{abelev} is preferable to further optimize the distinguishability between true jet and non-jet components in $\Delta\eta$. This work is in progress but it is beyond the scope of these proceedings.

%
%\clearpage
%
%%%%%%%%%%%%%%%%%%%%%%%%%%%%%%%%%%
\section{Discussion and Summary}

The momentum dependence of the amplitude and $\Delta\phi$ width was recently used in theory papers to distinguish contributions from initial bulk matter conditions (e.g. color glass condensate) and jet like correlations \cite{gavin,shuryak}. Fig.~\ref{figure:amp-phi} shows the relevant data.

The measured features (decrease in $\Delta\phi$ width and amplitude) are in general
agreement with predictions, although we should note that these models as of now only describe these two parameters of the correlation structures. No calculations regarding the $\Delta\eta$ width or the simultaneous description of the away-side structures have been attempted. Furthermore less quantitative alternative approaches, such as the evolution of mini-jets \cite{trainor2}  or the initial parton interaction models \cite{wong,Hwa} claim to describe the data as well. These models will have to be expanded to a level where one can actually perform direct comparisons between data and theory. For now the strong correlation of particles in $\Delta\eta$-$\Delta\phi$, even without requiring  a high p$_{T}$ trigger, remains an intriguing experimental fact which will be further explored, together with the analysis of other reported structures such as the conical emission of particles on the away-side \cite{jia,fuqiang}. It might well be that there is a common origin for all reported correlation structures and progress in modeling, such as the recent attempt to evolve the initial conditions in a hydrodynamical transport code \cite{takahashi}, will further constrain all possible explanations.

%%%%%%%%%%%%%%%%%%%%%%%%%%%%%%%%%%

\end{document}